\documentclass{article}
\usepackage{amsmath}
\usepackage{graphicx}
\usepackage{geometry}
\usepackage{subfig}
\usepackage{multirow}%
\usepackage{amsthm}%
\usepackage{mathrsfs}%
\usepackage[title]{appendix}%
\usepackage{xcolor}%
\usepackage{textcomp}%
\usepackage{manyfoot}%
\usepackage{upgreek}
\usepackage{booktabs}%
\usepackage{algorithm}%
\usepackage{algorithmicx}%
\usepackage{algpseudocode}%
\usepackage{listings}%

\title{\LARGE Systems-Level Analysis of Multisite Protein Phosphorylation: Mathematical Induction, Geometric Series, and Entropy}

\author{\large Iman Tavassoly$^{1,*,**}$, Adel Mehrpooya$^{2,3, *,**}$, Parsa Mirlohi$^{4}$, Zahra Abbaspourasadollah$^{5}$}

\theoremstyle{thmstyleone}%
\theoremstyle{thmstyletwo}%
\newtheorem{example}{Example}%

\theoremstyle{thmstylethree}%

\date{}  % no date

\begin{document}
\maketitle

\begin{center}
\small
1. QMed Insights, New York, NY, USA \\
2. School of Mathematical Sciences, Faculty of Science, Queensland University of Technology (QUT), 2 George Street, Brisbane, 4000, Queensland, Australia \\
3. Max Planck Queensland Centre for the Materials Science of Extracellular Matrices, Brisbane, Queensland, Australia \\
4. Department of Computer Sciences, Virginia Polytechnic Institute and State University, Blacksburg, VA, USA \\
5 Genetics, Bioinformatics and Computational Biology Program, Virginia Polytechnic Institute and State University, Blacksburg, VA 24060, USA \\
*First Co-Authors
** Corresponding authors: iman.tavassoly@qmedinsights.com, adel.mehrpooya@hdr.qut.edu.au;
\end{center}

\vspace{0.5cm}


\vspace{0.5cm}

\maketitle

\begin{abstract}
Multisite protein phosphorylation plays a pivotal role in regulating cellular signaling and decision-making processes. In this study, we focus on the mathematical underpinnings and informational aspects of sequential, distributive phosphorylation systems. We first provide rigorous steady-state solutions derived using geometric series arguments and formal mathematical induction, demonstrating that the distribution of phosphorylation states follows a geometric progression determined by the kinase-to-phosphatase activity ratio. We then extend the analysis with entropy-based insights, quantifying uncertainty in phosphorylation states and examining the mutual information between kinase activity and phosphorylation levels through a truncated Poisson model. These results highlight how phosphorylation dynamics introduce both structured patterns and inherent signal variability. By combining exact mathematical proofs with entropy analysis, this work clarifies key quantitative features of multisite phosphorylation from a systems-level perspective.

\end{abstract}

\section{Introduction and System Description}
Protein phosphorylation is one of the most versatile post-translational modifications that regulate cellular processes such as signal transduction, cell cycle control, metabolic regulation, and cell death \cite{Graves1999, Tavassoly20152}. Multisite phosphorylation, where a single substrate possesses multiple phosphorylation sites, enables cells to create fine-tuned control mechanisms, ultrasensitive responses, bistable switches, and memory storage \cite{Kapuy2009, Gunawardena2005, Tavassoly2015}.

Mathematical modeling provides a quantitative lens for investigating these dynamic processes, particularly using Ordinary Differential Equations (ODEs) that describe the time evolution of different phosphorylation states  \cite{Tavassoly2018, Tavassoly2023}. In this paper, we derive the steady-state distributions of phosphorylation states using two mathematical approaches: a classical geometric series argument and a formal mathematical induction proof. We explore the origins of these equations, explain the proof approaches, and connect the mathematical results to biological interpretations.

\begin{figure}[h]
\centering
\includegraphics[width=1\textwidth]{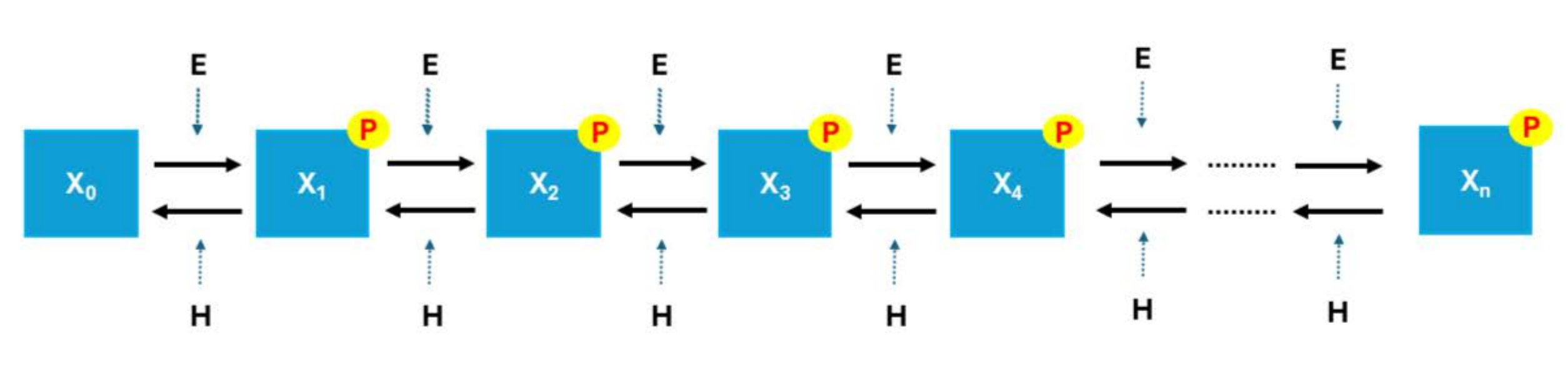}
\caption{Illustration of sequential, distributive multisite phosphorylation by kinase (E) and dephosphorylation by phosphatase (H).}
\label{fig:figure1}
\end{figure}

We examine a substrate protein $X$ that possesses $N$ phosphorylation sites, allowing it to exist in multiple biochemical states depending on how many of these sites are phosphorylated. The fully unphosphorylated form is indicated by $X_0$, while $X_i$ represents the form phosphorylated at specific sites $i$ (with $i$ ranging from 1 to $N$), and $X_N$ denotes the fully phosphorylated species. Biochemical reactions involve two key enzymes: the kinase, labeled $E$, which catalyzes the phosphorylation process, and the phosphatase, labeled $H$, which catalyzes the dephosphorylation steps.

Importantly, phosphorylation follows a sequential and ordered mechanism: kinase $E$ adds phosphate groups one at a time, following a strict site order, moving progressively from $X_0$ to $X_1$, $X_1$ to $X_2$, etc., until it reaches $X_N$. In contrast, phosphatase $H$ removes phosphate groups in reverse sequence, going from $X_N$ back to $X_{N-1}$, $X_{N-1}$ to $X_{N-2}$, and finally returning the substrate to $X_0$. This sequential and distributive nature means that each binding event by kinase or phosphatase modifies only one site before dissociation (Figure 1).

Assuming the system operates under binding-limited kinetics, where the rate-limiting step is the enzyme, substrate binding event, rather than the catalytic conversion, the steady-state dynamics can be described by a set of ordinary differential equations \cite{Kapuy2009}:
\begin{equation}
k_e E X_{i-1} = k_h H X_i, \quad i = 1, 2, \dots, N,
\end{equation}
where $k_e$ and $k_h$ are the catalytic rates of the kinase and phosphatase, respectively.

To simplify the analysis, we define the catalytic efficiency ratio:\begin{equation}
r = \frac{k_e E}{k_h H},
\end{equation}
which allows us to express the concentration of each phosphorylated species in terms of the unphosphorylated form:
\begin{equation}
X_i = r^i X_0.
\end{equation}The conservation of total protein, meaning the sum of all phosphorylation states equals the total substrate pool $X_T$ is written as:
\begin{equation}
X_T = \sum_{i=0}^N X_i = X_0 (1 + r + r^2 + \dots + r^N).
\end{equation}
These fundamental equations arise from the considerations of mass balance in the steady state, where the incoming flux into each state equals the outgoing flux. They form the backbone of the mathematical proofs we present in the following sections.

\section{Classical Geometric Series Proof}

We begin by deriving the steady-state concentrations using an explicit algebraic approach. At steady state, the flux into and out of each phosphorylated species must balance. For each phosphorylation step, the forward flux mediated by the kinase ($E$) matches the reverse flux mediated by the phosphatase ($H$). This yields the steady-state relationship:\begin{equation}
k_e E X_{i-1} = k_h H X_i, \quad i = 1, 2, \dots, N.
\end{equation}Rearranging, we express each $X_i$ in terms of its predecessor:
\begin{equation}
X_i = \frac{k_e E}{k_h H} X_{i-1} = r X_{i-1},
\end{equation}where we define $r = \frac{k_e E}{k_h H}$, the catalytic efficiency ratio. By recursive substitution, we obtain:
\begin{equation}
X_i = r^i X_0,
\end{equation}where $X_0$ is the fully unphosphorylated species.

Next, we apply the total protein conservation condition, recognizing that the total amount of substrate protein, $X_T$, is fixed:
\begin{equation}
X_T = \sum_{i=0}^N X_i = X_0 (1 + r + r^2 + \dots + r^N).
\end{equation} We recognize this as a finite geometric series:
\begin{equation}
1 + r + r^2 + \dots + r^N = \frac{1 - r^{N + 1}}{1 - r}, \quad r \neq 1.
\end{equation}Solving for $X_0$, we obtain:\begin{equation}
X_0 = X_T \cdot \frac{1 - r}{1 - r^{N + 1}}.
\end{equation}Substituting back, we find the concentration of each phosphorylated species:\begin{equation}
X_i = r^i X_0 = X_T \cdot \frac{(1 - r) r^i}{1 - r^{N + 1}}.
\end{equation}When $r = 1$, the geometric series simplifies:
\begin{equation}
1 + 1 + \dots + 1 = N + 1,
\end{equation}
yielding:
\begin{equation}
X_0 = \frac{X_T}{N + 1}, \quad X_i = X_0.
\end{equation}This derivation reveals that under steady-state and binding-limited conditions, the relative abundance of each phosphorylated state depends only on the ratio of kinase to phosphatase activity, not on their absolute concentrations. The geometric scaling encodes how enzymatic balance shapes signal propagation, threshold sensitivity, and dynamic range, making it a powerful predictive tool in systems biology.

\section{Mathematical Induction Proof}

To rigorously prove that the steady-state concentrations satisfy $X_i = r^i X_0$ for all $i = 1, 2, \dots, N$, we apply the method of mathematical induction \cite{Bather1994}. This formal approach ensures the generality and correctness of our result.
We aim to prove:
\begin{equation}
X_i = r^i X_0, \quad \text{for all} \ i = 1, 2, \dots, N,
\end{equation}
where $r = \frac{k_e E}{k_h H}$ and $X_0$ is the unphosphorylated concentration.

This result confirms that the steady-state concentrations follow a precise and predictable pattern — a geometric progression — across all phosphorylation states.

\subsection{Base Case ($i = 1$)}

First, we check the smallest nontrivial case. At steady state:\begin{equation}
k_e E X_0 = k_h H X_1.
\end{equation}Rearranging:
\begin{equation}
X_1 = \frac{k_e E}{k_h H} X_0 = r X_0.
\end{equation}Thus, the formula holds for $i = 1$.

\subsection{Induction Hypothesis}

Assume the relation: \begin{equation}
X_n = r^n X_0.
\end{equation}holds for an arbitrary index n; this serves as our inductive hypothesis.

\subsection{Inductive Step}

We now prove the statement holds for $n + 1$. Using the steady-state balance: \begin{equation}
k_e E X_n = k_h H X_{n + 1},
\end{equation}
we solve:
\begin{equation}
X_{n + 1} = \frac{k_e E}{k_h H} X_n = r X_n.
\end{equation}By the inductive hypothesis:
\begin{equation}
X_n = r^n X_0,
\end{equation}thus:
\begin{equation}
X_{n + 1} = r (r^n X_0) = r^{n + 1} X_0.
\end{equation}By mathematical induction, the formula:
\begin{equation}
X_i = r^i X_0,
\end{equation}
holds for all $i = 1, 2, \dots, N$.

This formal proof guarantees that the steady-state pattern is not an artifact of specific cases but a general result for all sequential, distributive multisite phosphorylation systems under the defined conditions. It provides a robust foundation for both theoretical understanding and future model extensions, ensuring that any added complexity builds upon a solid and proven mathematical base.

\section{Biological Significance}
The analytical proofs derived in this work reveal a fundamental property of multisite phosphorylation systems: at steady state, the distribution of phosphorylated species follows a geometric progression governed solely by the ratio of kinase to phosphatase activity. This mathematical insight captures the core mechanism underlying many well-characterized biological phenomena, including threshold responses, ultrasensitivity, and bistable switching observed in key signaling networks. Such quantitative clarity not only deepens our understanding of cellular signal processing but also enhances our ability to predict system behavior under perturbation, identify sensitive control points for pharmacological intervention, and design synthetic regulatory circuits that mimic natural control mechanisms. 
While the current work establishes a general and rigorous analytical framework under the simplifying assumptions of linear, binding-limited kinetics, it serves as only the first step toward modeling the rich complexity of biological regulation. Many cellular systems operate under nonlinear saturation regimes, exhibit cooperative binding, or assemble multienzyme complexes, leading to behaviors that cannot be captured by linear models alone. Additionally, stochastic fluctuations, particularly relevant in systems with low molecular counts, can introduce probabilistic switching and heterogeneous cellular responses that demand stochastic or hybrid modeling frameworks. Future extensions will generalize these steady-state proofs by incorporating nonlinear kinetics (e.g., Michaelis-Menten saturation), cooperative effects, feedback and feedforward motifs, and stochastic dynamics using tools such as perturbation analysis, bifurcation theory, and stochastic simulations (e.g., Gillespie algorithms). Integrating these advanced mathematical tools with computational modeling will not only refine our understanding of robustness and adaptability in biological systems but also support the rational design of therapeutic interventions and synthetic networks with engineered dynamic behaviors.

\section{Entropy and Information: Theoretic Perspectives on Multisite Phosphorylation}
A substrate protein with $N$ phosphorylation sites can exist in $N + 1$ discrete states, from unphosphorylated to fully phosphorylated. The transitions among these states, catalyzed by kinases and phosphatases, are inherently stochastic and probabilistic, particularly in the distributive, sequential model where only one phosphorylation event occurs at a time \cite{PP10}. This stochastic nature raises important questions: How reliably can the system transmit biochemical signals? How much uncertainty exists in the phosphorylation state over time? And to what extent can the phosphorylation state infer the upstream signal?

To address these questions, tools from information theory, notably Shannon entropy and mutual information, have been employed to quantify the uncertainty and information-processing capacity of phosphorylation systems \cite{PP00,PP05,PP07}. The concept of Measure-Theoretic Entropy has been applied to various systems with information transitions between different states, characterizing the uncertainty in those systems' states \cite{PP01,PP02,PP03,PP04,PP06}. It has been an effective method for measuring the uncertainty of random variables modeling biological systems including phosphorylation systems \cite{PP00,PP05,PP06,PP07}.

Mutual information has also been used effectively in multisite phosphorylation studies to quantify the amount of information the phosphorylation state conveys about the upstream kinase signal. This metric is particularly relevant in biological systems where the precision and fidelity of signal transduction are critical for cell fate decisions, metabolic regulation, and immune responses \cite{PP07,PP08,PP09}.

In this section, we analyze a three-site sequential, distributive phosphorylation system with symmetric transition rates, using a truncated Poisson model to describe the time evolution of phosphorylation probabilities. We compute the Shannon entropy as a dynamic indicator of system uncertainty and assess the mutual information between kinase activity and phosphorylation state, revealing how signal sharpness, noise, and system resolution vary with kinase strength and number of phosphorylation sites. Our results demonstrate multisite phosphorylation increases signaling complexity but transmits limited information due to overlapping states and noise, and high fidelity requires ultrasensitivity or cooperativity. Thus, information theory quantifies signaling precision and highlights constraints in biochemical information transmission.

Let $X$ be a protein with $N$ phosphorylation sites. Since in a sequential, distributive model, only one site is phosphorylated at a time, the order of phosphorylation matters, and the phosphorylation process evolves over time, we can assume that a substrate is in state $X_i$ with a probability $p_i(t)$ at a given time for $i = 0, 1, 2, \ldots, N$. Then the substrate phosphorylation entropy is:
\begin{equation}
	H_{\mathrm{SP}}(X) = - \sum_{i = 0}^{N} p_i(t) \log_2 p_i(t).
\end{equation}

The substrate phosphorylation entropy $H_{\mathrm{SP}}(X)$ characterizes the diversity and uncertainty in the system's phosphorylation states, and measures the information content associated with the probability distribution of the phosphorylation level of a randomly chosen protein molecule $X$. It provides a time-resolved picture of how the system gains or loses informational order, for instance, converging toward a steady-state, or sharpening a signal. It also quantifies the diversity of phosphorylation states in a population of substrates. If $H_{\mathrm{SP}}(X)$ is low, the signal is strong with high certainty indicating mostly fully phosphorylated state, whereas a high value of $H_{\mathrm{SP}}(X)$ reflects a weak or noisy signal indicating greater variability of phosphorylation.

We can quantify how much information about the signals corresponding to kinase activity is transmitted to the phosphorylation state using mutual information defined by:
\begin{equation}
	I(K;X) = H_{\mathrm{SP}}(K) - H_{\mathrm{SP}}(K|X), 
\end{equation}
where $K$ is the input signal arisen from the kinase concentration, $X$ is the phosphorylation state, and $H_{\mathrm{SP}}(K|X)$ is the conditional entropy indicating uncertainty in input given observed output.

In physical terms, analogous to a noisy electronic communication channel, the kinase functions as the signal sender, while the phosphorylation state represents the received message, potentially distorted due to stochastic biochemical transitions. The mutual information $I(K;X)$ quantifies the reduction in uncertainty about the phosphorylation state $X$ given knowledge of the kinase activity $K$. In other words, it measures the extent to which kinase activity is reliably transmitted through the biochemical network and encoded in phosphorylation patterns. A higher value of $I(K;X)$ indicates greater signal fidelity and more distinguishable phosphorylation states.

\begin{example}
	{\textbf{3-site Sequential Phosphorylation:}} 
	Suppose a substrate has four sites, phosphorylation rate $K_{\mathrm(p)}$ and dephosphorylation rate $K_{\mathrm(d)}$ are equal, $K_{\mathrm(p)} = K_{\mathrm(d)} = K$. We assume that the phosphorylation system behaves like a Poisson process under symmetric rates truncated at $n = 3$ since we have only 4 states. The initial probabilities are $p_0(0) = 1$, $p_1(0) = p_2(0) = p_3(0) = 0$, and the current state probabilities are:
	\begin{equation}
		p_n(t) = \dfrac{3^nK^nt^n}{n!}\exp(-Kt), \quad \text{ for } n = 0, 1, 2, 3.
	\end{equation}	
%	 $p_0(t) = \exp(-Kt)$, $p_1(t) = 3Kt\exp(-Kt)$, $p_2(t) = \frac{9}{2}K^2t^2\exp(-Kt)$, and $p_3(t) = \frac{27}{6}K^3t^3\exp(-Kt)$.
	\begin{figure}[h!]%
		\centering
%		\hspace{-5cm}
		\includegraphics[width=0.8\textwidth]{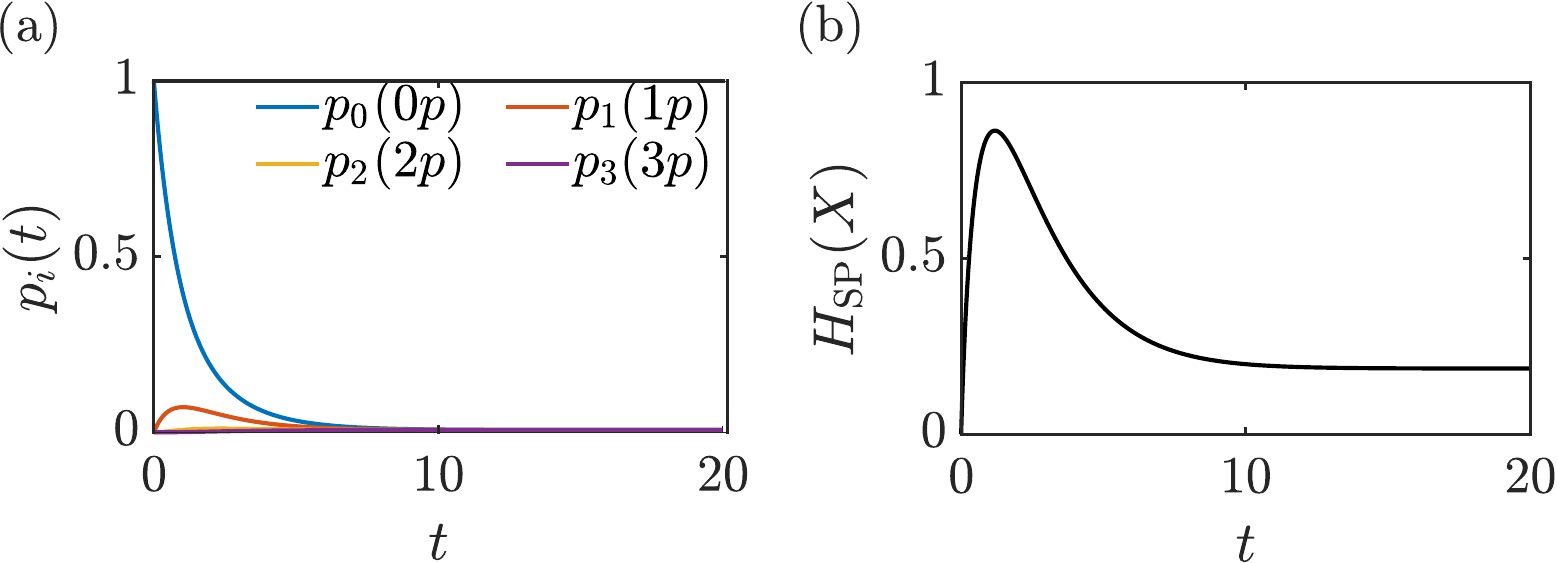}
		
		\caption{(a) Time-dependent probabilities $p_n(t)$ for a substrate protein with three phosphorylation sites ($n = 0, 1, 2, 3$) undergoing sequential, distributive phosphorylation and dephosphorylation at equal rates $K$. The curves represent the truncated Poisson dynamics of phosphorylation states over time. (b) Shannon entropy $H_{\mathrm{SP}}(X)$ of the phosphorylation state distribution over time for a substrate with three phosphorylation sites undergoing sequential, distributive (de)phosphorylation at equal rates $K = 1$.}
		\label{f2}
	\end{figure}
	Figure \ref{f2} illustrates the time-dependent behavior of the probabilities $p_i(t)$ for $i = 1, 2, 3, 4$, as well as the corresponding entropy evolution. The entropy curve obtained from the probabilities $p_i(t)$ illustrates the rate at which the system diverges from a known initial state and becomes more uncertain, while also indicating how much information is preserved or dissipated over time. Consequently, entropy serves as a diagnostic tool for assessing the fidelity, molecular noise, and dynamic characteristics of the intricate biological system of multisite phosphorylation. In particular, low entropy denotes a precise signal, implying that all proteins are fully phosphorylated, whereas high entropy signifies an ambiguous signal, indicating a difficult-to-discern state.
	
	The peak entropy corresponds to the maximum uncertainty in the state and may occur at time points that coincide with transition phases in signaling cascades. This highlights when a pathway is most adaptable. The entropy plateau is indicative of the system's steady state, suggesting that the signaling system has established a stable response pattern. This information facilitates the assessment of the information-processing capacity of the phosphorylation system. Specifically, a system exhibiting sharpened responses would exhibit low steady-state entropy, whereas a system featuring graded responses would accommodate higher entropy levels.		
	\begin{figure}[h!]%
		\centering
		\includegraphics[width=0.8\textwidth]{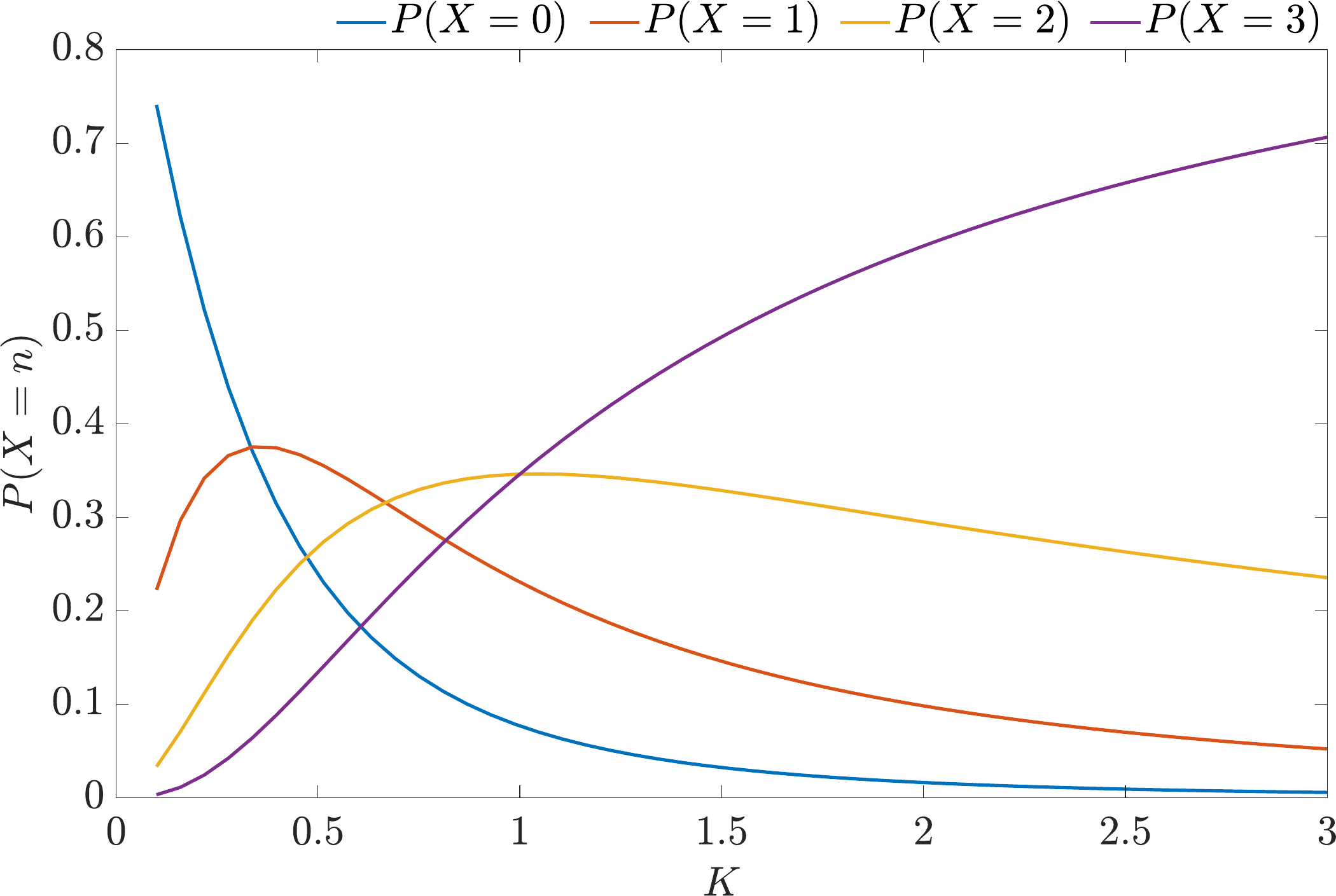}
		
		\caption{Steady-state probabilities of a substrate protein existing in each phosphorylation state $X = 0, 1, 2, 3$ as a function of kinase activity $K$, assuming a truncated Poisson distribution for state probabilities with symmetric transition rates. As kinase activity increases, the system transitions from predominantly unphosphorylated states ($X = 0$) to highly phosphorylated states ($X = 3$). The sigmoidal-like transitions reflect the probabilistic nature of phosphorylation events.}
		\label{f3}
	\end{figure}

	Figure \ref{f3} demonstrates the change in probability of specific phosphorylation states with increasing kinase activity. The $P(X=0)$ (red) curve indicates that the protein remains predominantly unphosphorylated under low kinase activity. The $P(X=1)$ (green) curve peaks at intermediate kinase activity ($K \sim 0.5$), suggesting a preference for single phosphorylation events. The $P(X=2)$ (blue) curve dominates at moderate kinase activity, implying an intermediate phosphorylation level. Lastly, the $P(X=3)$ (black) curve represents the fully phosphorylated substrate under high kinase activity. These curves undergo sigmoidal transitions, illustrating how phosphorylation patterns vary with kinase signals.
	\begin{figure}[h!]%
		\centering
		\includegraphics[width=1.0\textwidth]{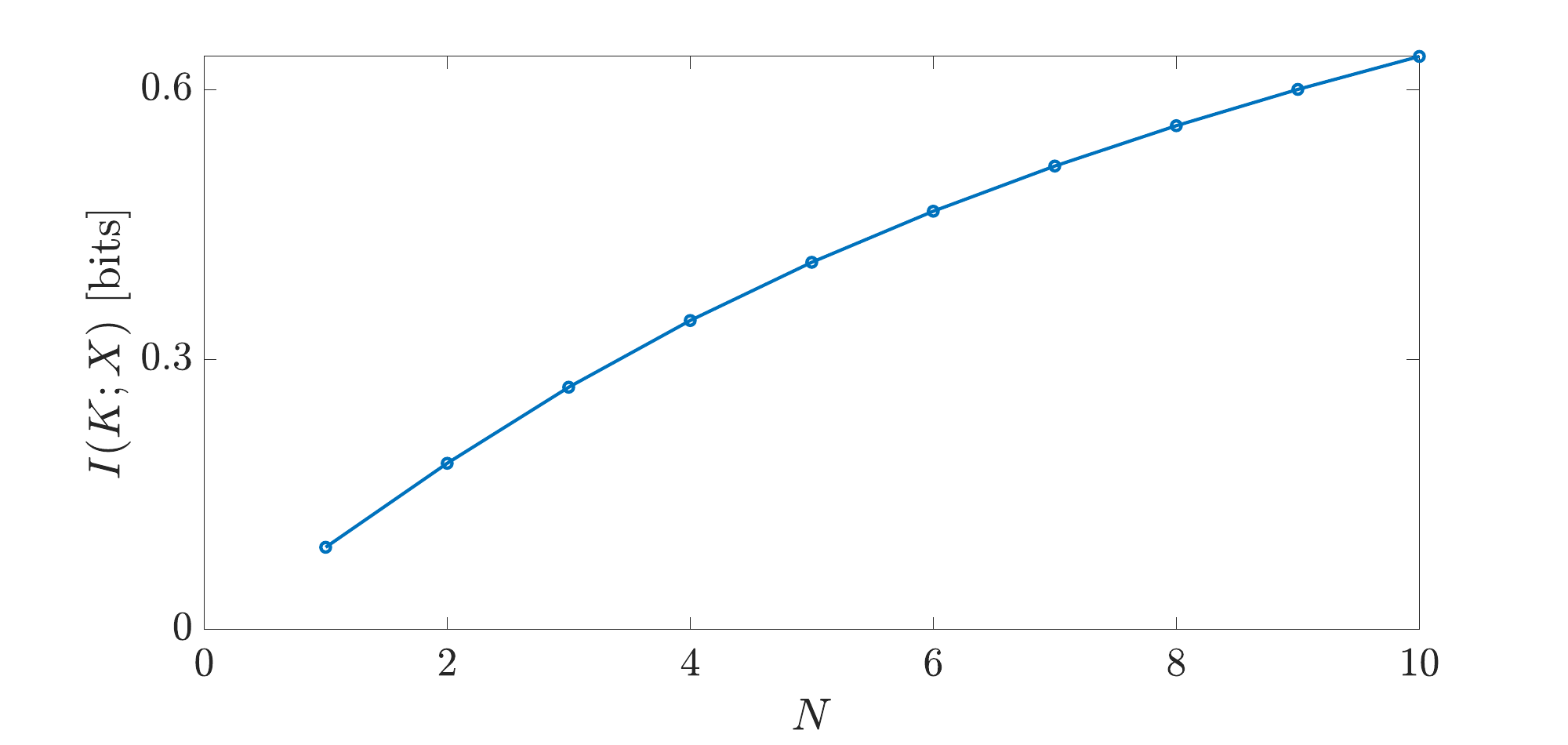}
		
		\caption{Mutual information versus the number of phosphorylation sites for $N = 1, 2, \ldots, N_{\mathrm{max}}$, assuming a truncated Poisson distribution for state probabilities with symmetric transition rates. As the number of phosphorylation sites increases, the mutual information $I(K;X)$ increases.}
		\label{f4}
	\end{figure}	
	The mutual information is calculated as $I(K;X) = 0.2929\,$bits, implying that the kinase signal $K$ transmits only $\sim 0.29\,$bits of information to the phosphorylation state $X$. This is substantially lower than the maximum possible value ($\log_2(4) = 2\,$bits) for four states, due to ambiguous state distributions, probabilistic transitions, and the absence of ultrasensitivity or cooperativity.
	
	Biologically, this limited information transmission restricts downstream proteins' ability to accurately interpret signals. Consequently, phosphorylation states do not completely resolve kinase activity, resulting in significant ambiguity. With four phosphorylation states, only $\sim 0.293\,$bits of signal are transmitted, indicating a constrained signaling precision.
	
	Figure~\ref{f4} shows the mutual information versus the number of phosphorylation sites, where the number of phosphorylation sites is $N = 1, 2, \ldots, N_{\mathrm{max}}$. For each $N$, mutual information $I(K;X)$ is computed using a truncated Poisson model over kinase activity values. The figure reveals a monotonic increase in $I(K;X)$ as $N$ increases; however, the growth is sublinear due to the increasing overlap and noise. It is evident that full resolution of kinase activity remains limited even with more states. This can be used to analyze the information capacity of multisite phosphorylation systems. Informationally, the capacity of a biochemical system to transmit input signals increases with the complexity of the system but not linearly. Biologically, more phosphorylation states may not yield proportionally more actionable signals unless the system also features ultrasensitivity, cooperativity, or state-dependent functions.	 
	\begin{figure}[h]%
		\centering
		\includegraphics[width=0.8\textwidth]{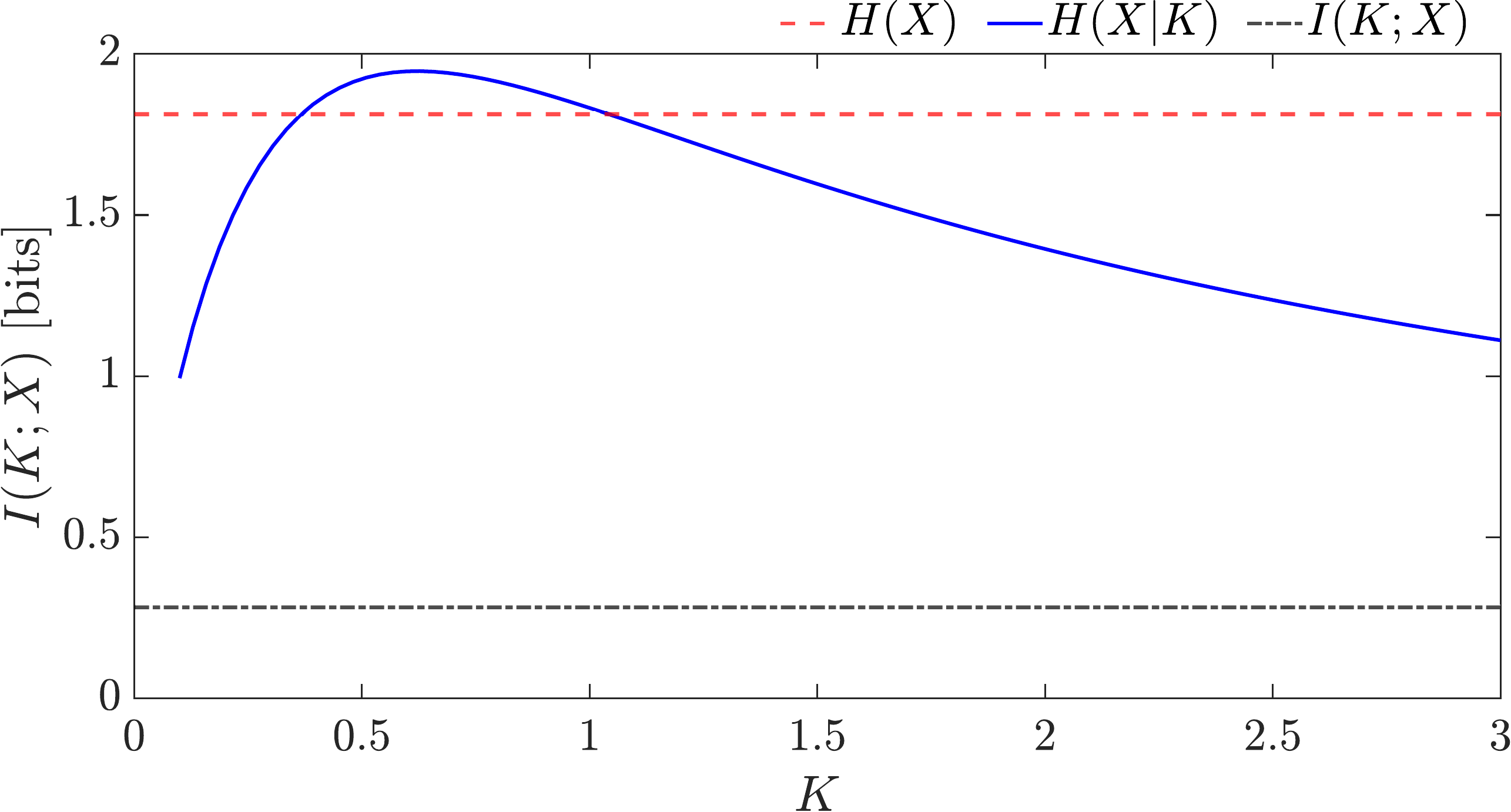}
		
		\caption{Mutual information $I(K;X)$ between kinase activity $K$ and the phosphorylation state $X$ for a substrate with $N = 3$ phosphorylation sites. The system follows a truncated Poisson distribution under symmetric phosphorylation and dephosphorylation rates. Mutual information quantifies the fidelity with which the substrate’s phosphorylation level encodes the kinase input. The curve shows that $I(K;X)$ reaches a maximum in an intermediate regime of kinase activity, where variability in phosphorylation states is high yet distinguishable. At low or high $K$, saturation or minimal activity reduces the system's ability to convey information, reflecting biological signal compression or ambiguity.}
		\label{f5}
	\end{figure}
	
	Figure~\ref{f5} displays the entropy $H(X)$ and mutual information $I(K;X)$ for multisite phosphorylation with symmetric rates based on the truncated Poisson distribution. The blue line represents conditional entropy $H(X | K=k)$ as kinase activity varies, indicating the uncertainty of state $X$ at each $K$. The red dashed line is marginal entropy $H(X)$, indicating uncertainty in the phosphorylation state when the kinase is not known. The black dashed-dotted line shows the mutual information $I(K;X)$, indicating the amount of information $X$ conveys about kinase activity $K$. As shown, $H(X) > H(X | K)$, indicating that the knowledge of $K$ reduces uncertainty about $X$. The fact that $I(K;X)\approx 0.293\,$ bits confirms the limited transmission of information due to overlap in $P(X | K)$.

	In general, these findings emphasize the impact of kinase activity and phosphorylation states on signaling precision and accuracy within biochemical systems.This entropy-based analysis offers a complementary perspective to the deterministic proofs presented earlier in this paper. By incorporating Shannon entropy and mutual information into the study of multisite phosphorylation, we quantified how signal precision is intrinsically limited by noise, overlapping states, and stochastic transitions. Using a truncated Poisson model, we characterized the dynamic distribution of phosphorylation states and assessed their information-processing capacity, showing that, while additional phosphorylation sites increase complexity, they do not necessarily enhance the fidelity of signal transmission. Overall, this entropy framework provides a systems-level probabilistic extension of classical steady-state analysis, revealing how multisite phosphorylation both encodes and constrains biochemical signals.
\end{example}

%\clearpage
\section{Discussion and Conclusions}
This study combines rigorous mathematical proofs with information-theoretic analysis to characterize the behavior of multisite protein phosphorylation systems. Using two complementary approaches, geometric series derivation and mathematical induction, we have established general steady-state solutions for sequential, distributive phosphorylation, showing that the distribution of phosphorylation states forms a geometric progression determined solely by the ratio of kinase to phosphatase activity. This analytical framework clarifies the underlying mathematical structure of phosphorylation dynamics and provides a foundation for further exploration of ultrasensitivity, threshold responses, and bistable behavior in signaling networks.

Building on this foundation, we introduced entropy and mutual information as measures of uncertainty and signal fidelity in multisite phosphorylation. Through a truncated Poisson model, we quantified how phosphorylation states encode kinase activity and how noise and overlapping states limit the information transmission capacity. Our results indicate that, while increasing the number of phosphorylation sites enhances the complexity of signaling patterns, it does not necessarily lead to proportionally higher information transfer. High signal fidelity requires additional mechanisms such as ultrasensitivity, cooperativity, or allosteric regulation.

Together, these mathematical and entropy-based analyses provide complementary perspectives: the proofs capture the deterministic structure of steady-state distributions, while the entropy metrics reveal probabilistic constraints and noise effects inherent to biochemical communication. Future extensions of this work could explore hybrid models that incorporate both nonlinear kinetics (e.g. Michaelis-Menten saturation) and stochastic simulations (e.g. Gillespie algorithms) to better capture real-world cellular environments. By bridging exact analytical results with information-theoretic insights, this study contributes to a systems-level understanding of how multisite phosphorylation balances precision, robustness, and adaptability in cell signaling.

%\clearpage

\end{document}